\documentclass[twocolumn]{aastex61}

\received{}
\revised{}
\accepted{}
\submitjournal{The Astronomical Journal}

\shorttitle{Masers in low-luminosity objects}
\shortauthors{G\'omez et al.}

\begin{document}

\title{A search for water maser emission from brown dwarfs and low-luminosity young stellar objects}

\correspondingauthor{Jos\'e F. G\'omez}
\email{jfg@iaa.es}

\author[0000-0002-7065-542X]{Jos\'e F. G\'omez}
\altaffiliation{On sabbatical leave at Laboratoire Lagrange, Universit\'e de Nice Sophia-Antipolis, CNRS, Observatoire de la C\^ote d'Azur, F-06304 Nice, France}
\affiliation{Instituto de Astrof\'{\i}sica de Andaluc\'{\i}a, CSIC, Glorieta de la Astronom\'{\i}a s/n, 18008 Granada, Spain}

\author[0000-0002-9569-9234]{Aina Palau}
\affiliation{Instituto de Radioastronom\'ia y Astrof\'isica, UNAM, P.O. Box 3-72, 58090, Morelia, Michoac\'an, M\'exico}

\author{Lucero Uscanga}
\affiliation{Departamento de Astronom\'{\i}a, Universidad de Guanajuato, A.P. 144, 36000 Guanajuato, Gto., M\'exico}

\author{Guillermo Manjarrez}
\affiliation{Instituto de Astrof\'{\i}sica de Andaluc\'{\i}a, CSIC, Glorieta de la Astronom\'{\i}a s/n, 18008 Granada, Spain}

\author[0000-0002-5971-9242]{David Barrado}
\affiliation{Centro de Astrobiolog\'{\i}a, INTA-CSIC, PO BOX 28692, ESAC Campus, E-208691 Villanueva de la Ca\~nada, Madrid, Spain}



\begin{abstract}

We present a survey for water maser emission toward a sample of 44 low-luminosity young objects, comprising (proto-)brown dwarfs, first hydrostatic cores (FHCs), and other young stellar objects (YSOs) with bolometric luminosities lower than 0.4 L$_\odot$. Water maser emission is a good tracer of energetic processes, such as mass-loss and/or accretion, and is a useful tool to study this processes with very high angular resolution. This type of emission has been confirmed in objects with L$_{\rm bol}\ga 1$ L$_\odot$. Objects with lower luminosities also undergo mass-loss and accretion, and thus, are prospective sites of maser emission. Our sensitive single-dish observations provided a single detection when pointing toward the FHC L1448 IRS 2E. However, follow-up interferometric observations showed water maser emission associated with the nearby YSO L1448 IRS 2 { (a Class 0 protostar of L$_{\rm bol}\simeq 3.6-5.3$ L$_\odot$)}, and did not find any emission toward L1448 IRS 2E. The upper limits for water maser emission determined by our observations are one order of magnitude lower than expected from the correlation between water maser luminosities and bolometric luminosities found for YSOs. This suggests that this correlation does not hold at the lower end of the (sub)stellar mass spectrum. Possible reasons are that the slope of this correlation is steeper at L$_{\rm bol}\le 1$ L$_\odot$, or that there is an absolute luminosity threshold below which water maser emission cannot be produced. Alternatively, if the correlation still stands at low luminosity, the detection rates of masers would be significantly lower than the values obtained in higher-luminosity Class 0 protostars.

\end{abstract}




\section{Introduction}

Water maser emission is present in different astrophysical environments, such as young stellar objects (YSOs), evolved stars, and active galactic nuclei \citep{eli92,lo05}. The most widely studied water maser transition is the one at 22 GHz, which has proved to be a very powerful tool to study these environments, since interferometric observations (in particular using Very Long Baseline Interferometry, VLBI) allow us to reach angular resolutions { better than} 1 milliarcsec \citep[e.g.,][]{tor01,tor14,yun11}. Thus, we can study processes and structures in these sources with a level of detail unsurpassed by any other technique in Astronomy.

Focusing on the case of water maser emission toward YSOs, we know that it is intense and widespread over a wide range of stellar masses {\citep[e.g.,][]{cla96,for99,fur03,bre10}}. Interferometric observations reveal a dichotomy in the distribution of water masers in YSOs:
while in some sources this type of emission traces collimated jets, in others it traces protoplanetary disks \citep[e.g.,][]{tor97,tor98}. This is important, since the presence of disks and collimated outflows are key ingredients in the
formation of stars via accretion.
In low-mass ($\la 4$ M$_\odot$) YSOs, water maser emission
 is consistently found associated with sources in the earliest evolutionary stages. Among objects in the classical evolutionary classes defined by \cite{lad87} and \cite{and93},  \cite{fur01} estimated detection rates of $\sim 40$\% in Class 0 objects, 4\% in Class I, and 0\% in Class II ones. This is understandable,
because water masers are thought to be collisionally pumped in post-shock
regions \citep{eli89,yat97}. The youngest YSOs seem to be the ones with more powerful collimated mass outflows \citep{bon96, dio10, pod12, wat16}, which can
create shocked regions as they move through the ambient molecular cloud. They also undergo intense accretion \citep{hen97}, which could pump maser emission in disks. 

Collimated outflows and accretion processes are present in YSOs with extremely low mass, and even in substellar objects. Spitzer data revealed the presence of a new class of objects \citep{you04,kau05}  named ``Very Low Luminosity Objects'' (VeLLO). These are objects embedded in dense cores, with internal luminosities $< 0.1$~L$_\odot$. Some of these objects show Herbig-Haro
{ (HH) emission} \citep{ste07b,com06} and bipolar
molecular outflows \citep{and99,bus07}, which are classical signposts of mass-loss
processes in higher-mass
YSOs. Moreover, even for brown dwarfs (BDs) there is mounting evidence of the presence of accretion
\citep[e.g.,][]{muz00,com00,fer01,bar03,luh03,alc14}, outflows \citep{whe05,whe07,pal14,mor15}, and disks \citep{nat01,bar07,sch08,dow15,tes16}. 
This evidence suggests that the formation of VeLLOs and BDs takes place in a similar way to solar-type stellar objects. Then, we see that the structures and processes that give rise to water maser emission (accretion/outflows) are also present in these very low mass objects, so it is possible that they also harbor water masers. The difference may be that the water maser emission could be weaker, as it happens when we compare low- and
high-mass
YSOs \citep{cla96,fur01}. The  energy output in mass-loss processes decreases with stellar luminosity \citep{ang95}, so there may be a mass limit below which maser emission cannot be produced. So far, the lowest-luminosity YSOs reported to possibly harbor a water maser is GF 9-2 \citep[$\simeq 0.3$ L$_\odot$;][]{fur03}. Extending this to lower luminosity objects would allow us to study mass-loss processes and accretion disks at the low luminosity end of the mass spectrum.

Another important category of low-luminosity
objects are first hydrostatic cores \citep[FHCs;][]{lar69}. These represent
the first stage of the evolution of protostars, immediately before dissociation of molecular hydrogen and the subsequent formation of a Class 0 protostar. Some of these FHCs show evidence for collimated outflows
\citep[e.g.,][]{che10}, and therefore, they might also harbor water masers. Given that a large fraction of Class 0 protostars are water-maser
emitters, a search toward the earlier FHC phase would establish the starting point of maser emission in low-mass YSOs.

In this paper we present a search for water maser emission toward a sample of low luminosity YSOs and substellar objects, to investigate whether their mass-loss processes are energetic enough to power water maser emission. As a longer term goal, the identification of water-maser emitters among these objects would allow us to design VLBI observations to study mass-loss processes in very low luminosity objects at the smallest possible spatial scales.

%
%

\section{Source sample}

We compiled 44 sources with declination $\geq -20^\circ$, reported in the literature with $L_{\rm bol}<0.4$ L$_\odot$, and/or $M_* <0.15$ M$_\odot$, with the main requirement that they show evidence of outflow/accretion and/or signs of youth (e.g., excess emission at millimeter wavelengths).

The sample of 44 sources comprises four groups: i) 20 young BDs; ii) 6 VeLLOs; iii) 7 FHCs; iv) 11 low luminosity objects (LLOs) with $0.1<L_{\rm bol}/{\rm L}_\odot <0.4$ and/or $0.075<M_*/{\rm M}_\odot<0.15$, not included in any of the groups above. 

BDs with signs of youth were taken from different samples: young BDs (most of them Class II objects) detected at 1.3 mm by \cite{kle03} and \cite{sch06}; the three young substellar objects classified by \cite{whi04} in a spectroscopic optical study of Class I young stellar objects in Taurus-Auriga;  Class I BD candidates proposed by 
\cite{pal12} and \cite{mor15} using submillimeter and centimeter observations; and Class 0 proto-BD presented in \cite{pal14}. Regarding VeLLOs and FHCs, we took sources from the compilation of \cite{pal14}, as well as the globule CB 130, which host several FHC candidates \citep{kim11}.

{ The rest of LLOs were selected from the} sources in the \cite{muz03} 
 sample with stellar masses $<0.15$ M$_\odot$, and estimated ages $<1$~Myr, with the exception of CIDA\,14, which is reported to have $\sim 0.17$ M$_\odot$ by \cite{bri99}. It is interesting to note that three out of the five selected sources from \cite{muz03} present signs of outflow activity from [OI] lines. We also included individual targets which were confirmed to be deeply embedded and of very low mass, such as the detections by \cite{kle03}, L1415-IRS, { a highly variable source (FU Ori-like object) with only 0.13 L$_\odot$, which is also driving HH emission} \citep[][2007b]{ste07a}, the triple system IRAS\,04325+2402AB/C, first reported by \cite{har99}, where one of the components could be substellar and driving an outflow \citep{sch08,sch10},  and IRAS\,04166+2706, a 0.4 L$_\odot$ source driving an extremely high-velocity and highly collimated outflow 
 \citep[e.g.,][]{taf04,san09,wan14}. We also included two low-mass dense cores reported by 
 \cite{cod97}, associated with infrared (IRAS) or submillimeter sources \citep{dif08}.

\section{Observations}

\subsection{Effelsberg}

Single-dish observations were carried out with the 100-m Effelsberg radio telescope of the Max-Planck-Institut f\"ur Radioastronomie, between 2015 September 29 and 2015 October 3 (observational project 14-14). We used the S14mm receiver, and tuned it to observe the $6_{16}-5_{25}$ transition of the H$_2$O molecule (rest frequency 22235.08 MHz). The full width at half maximum (FWHM) of the telescope beam at this frequency is $39''$. The backend was the { Fast Fourier Transform Spectrometer (XFFT)}, with a bandwidth of 100 MHz (total velocity coverage of 1348 km s$^{-1}$) sampled over 32768 channels, providing an effective spectral resolution of 3.5 kHz (0.047 km s$^{-1}$). For all sources, the spectral bandpass was centered at a velocity with respect to the local standard of rest (kinematical definition, $V_{\rm LSRK}$) of zero. We checked the pointing accuracy of the telescope at least once every hour using nearby calibrators. The rms pointing accuracy was better than $2''$. We observed in position switching mode, alternating on- and off-source scans of 60 seconds each. { Each source was typically observed for a total time (on+off) of 30 minutes, although integrations were longer in some cases (e.g., for targets at low elevation)}. Data were processed with the Continuum and Line Analysis Single-dish Software (CLASS), which is part of the  GILDAS\footnote{http://www.iram.fr/IRAMFR/GILDAS} package of the Institut de Radioastronomie Milim\'etrique. The spectra were corrected by the elevation-dependent gain of the telescope and by atmospheric opacity, and were Hanning-smoothed to obtain a final spectral resolution of 0.33 km s$^{-1}$.  Table \ref{tab:observed} lists the observed positions and the resulting rms of the spectra, which is of the order of 20 mJy in all cases. Note that some sources were observed in different { epochs}. Given the usually high variability of water masers, we did not average the data of different days, but present the results of the observations of each day individually. Moreover, although our target list comprised 44 sources, only 42 positions were observed, since in the B1 and CB130 fields, the Effelsberg beam included two sources.

\startlongtable
\begin{deluxetable*}{lllcllcl}
	\tablecaption{Observed { positions}\label{tab:observed}}
	\tablewidth{0pt}
	\tablehead{
	\colhead{ Target name} & \colhead{R.A. (J2000)} & \colhead{Dec. (J2000)} & 
	\colhead{rms} & \colhead{Date} & \colhead{Type\tablenotemark{a}} & \colhead{$L_\mathrm{bol}$} & \colhead{References}\\
\colhead{}			&	\colhead{}			&	\colhead{}			& \colhead{(mJy)} & \colhead{} &\colhead{}	&	\colhead{($L_\odot$)} & \colhead{}}
		\startdata
L1451-mm 			& 03:25:10.25 & +30:23:55.0 & 21 & 2015-SEP-29	& FHC & 0.05	&1\\ 
L1448 IRS 2E	 		& 03:25:25.52 & +30:45:02.5 & 16 & 2015-SEP-29 & FHC	&$<0.1$	&2\\ 
						&			&			& 14 & 2015-OCT-03\\
Per-Bolo 58	& 03:29:25.40 & +31:28:15.0 & 18 & 2015-OCT-01	& FHC	&0.18 &3,4\\ 
B1-b field\tablenotemark{b} & 03:33:21.25 & +31:07:35.0 & 19 & 2015-SEP-30 & FHC	&0.15--0.31	&5,6,7 \\ 
IC 348-SMM2E & 03:43:57.73 & +32:03:10.1 & 18 & 2015-SEP-30	& BD-0	&0.10	&8\\ 
							&			&				& 20 & 2015-OCT-03\\
IC 348-173		& 03:44:09.98 & +32:04:05.4 & 18 & 2015-OCT-01 & LL0-I	&0.11	&9\\
IC 348-613 		& 03:44:26.90 & +32:09:25.0 & 19 & 2015-SEP-30	& BD-I &0.002	&10,11,12\\
IC 348-205 		& 03:44:29.54 & +32:00:53.2 & 18 & 2015-OCT-02 & LLO-I	&0.08	&9,13\\
IC 348-382 		& 03:44:30.78 & +32:02:42.9 & 27 & 2015-OCT-03 & LLO-I 	&0.012	&9,13\\
IC 348-165		& 03:44:35.43 & +32:08:54.4 & 26 & 2015-OCT-03 & LLO-I 	&0.10	&9,13\\
SSTB213 J041726.38+273920.0 & 04:17:26.38 & +27:39:20.0 & 18 & 2015-OCT-01	& BD-I &$<0.002$	&14,15\\
SSTB213 J041740.32+282415.5	& 04:17:40.32 & +28:24:15.5 & 17 & 2015-OCT-01	& BD-I &$<0.003$	&14,15\\
SSTB213 J041757.75+274105.5 & 04:17:57.77 & +27:41:05.0 & 17 & 2015-OCT-01 & BD-I &$<0.004$	&14,15,16\\
SSTB213 J041828.08+274910.9		& 04:18:28.08 & +27:49:10.9 & 18 & 2015-OCT-01& BD-I 	&$<0.001$	&14,15\\
SSTB213 J041836.33+271442.2		& 04:18:36.33 & +27:14:42.2 & 18 & 2015-OCT-01& BD-I 	&$<0.003$	&14,15\\
SSTB213 J041847.84+274055.3		& 04:18:47.84 & +27:40:55.3 & 17 & 2015-OCT-01	& BD-I &$<0.004$	&14,15\\
KPNO-Tau 2& 04:18:51.16 & +28:14:33.2 & 17 & 2015-OCT-01	& BD-II & $-$	&13,17,18\\
IRAS 04158+2805 			& 04:18:58.14 & +28:12:23.5 & 21 & 2015-SEP-30	& BD-I 	&0.4	&19,20,21\\
SSTB213 J041913.10+274726.0		& 04:19:13.10 & +27:47:26.0 & 16 & 2015-OCT-01& BD-I 	&$<0.002$	&14,15\\
SSTB213 J041938.77+282340.7		& 04:19:38.77 & +28:23:40.7 & 19 & 2015-OCT-02& BD-I 	&$<0.006$	&14,15\\
IRAS 04166+2706 		& 04:19:42.50 & +27:13:40.0 & 27 & 2015-OCT-02	&LLO-0 &$<0.39$	&22,23,24,25\\
L1521D 				& 04:21:03.55 & +27:02:48.4 & 28 & 2015-OCT-03	&LLO &0.3	&26\\
SSTB213 J042019.20+280610.3				& 04:20:19.20 & +28:06:10.3 & 18 & 2015-OCT-02	& BD-I &$<0.002$	&14,15 \\
SSTB213 J042118.43+280640.8		& 04:21:18.43 & +28:06:40.8 & 18 & 2015-OCT-02	& BD-I &$<0.002$	&14,15 \\
IRAM 04191+1522			& 04:21:56.91 & +15:29:45.9 & 20 & 2015-SEP-30	& VeLLO-0 &0.15	&27,28,29,30\\ 
IRAS 04248+2612 		& 04:27:57.33 & +26:19:18.1 & 19 & 2015-SEP-30	& BD-I 	&0.4	&20\\
L1521F-IRS 			& 04:28:38.95 & +26:51:35.1 & 19 & 2015-OCT-02	& VeLLO-0/I &$<0.07$	&30,31\\
MHO 5 		& 04:32:16.00 & +18:12:46.0 & 18 & 2015-SEP-30	&LLO-I &0.066	&9,13,32\\
IRAS 04325+2402 A/B		& 04:35:35.37 & +24:08:19.5 & 18 & 2015-OCT-02	&LLO-I &0.7	&33,34,35,36,37\\
2MASS J04381486+2611399 & 04:38:14.86 & +26:11:39.9 & 17 & 2015-SEP-30	& BD-II &$-$	&17,18,38\\
2MASS J04390396+2544264 & 04:39:03.96 & +25:44:26.4 & 19 & 2015-SEP-30	& BD-II &0.019	&17,18,32\\
CFHT-BD-Tau 4 & 04:39:47.30 & +26:01:39.0 & 19 & 2015-OCT-02 &LLO-II &$-$	&12,13,18,39,40\\
L1415-IRS & 04:41:35.87 & +54:19:11.7 & 18 & 2015-OCT-01 &LLO-I	&0.13	&41,42\\ 
2MASS J04414825+2534304 & 04:41:48.25 & +25:34:30.5 & 19 & 2015-SEP-30	& BD-II &0.009	&17,18,32\\
L1507A 				& 04:42:38.59 & +29:43:55.3 & 18 & 2015-OCT-02	&LLO&$<0.3$	&26 \\
2MASS J04442713+2512164 & 04:44:27.13 & +25:12:16.4 & 19 & 2015-SEP-30		& BD-II &0.028	&18,43\\
IRAS 04489+3042 		& 04:52:06.68 & +30:47:17.5 & 17 & 2015-OCT-01	& BD-I 	&0.3	&19,20\\
CB130-1 field\tablenotemark{c}  & 18:16:16.90 & $-02$:32:38.0 & 25 & 2015-SEP-30	& FHC&0.071-0.23	&44\\ 
						&			&					& 24 & 2015-OCT-01\\
L328-IRS				& 18:17:00.40 & $-18$:01:52.0 & 24 & 2015-SEP-30	& VeLLO-0 &0.14	& 45,46\\
						&				&				& 19 & 2015-OCT-01\\
L673-7-IRS & 19:21:34.80 & +11:21:23.0 & 15 & 2015-SEP-29	& VeLLO-0 	&0.18	&47,48,49\\ 
L1148-IRS  				& 20:40:56.59 & +67:23:05.3 & 14 & 2015-SEP-30	& VeLLO-0&0.12	&50\\
L1014-IRS			& 21:24:07.55 & +49:59:09.0  & 21 & 2015-SEP-30	& VeLLO-I/II	&0.15	&51,52\\
						&				&				& 17 & 2015-OCT-01\\
		\enddata
		\tablenotetext{a}{Type of object. BD: (proto-)brown dwarf. FHC: First hydrostatic core. VeLLO: Very low luminosity object. LLO: Other low-luminosity YSO. The evolutionary class of the object is indicated by 0, I, II (for Class 0, I, and II objects, respectively).}
		\tablenotetext{b}{The B1-b field includes 2 candidate FHCs within the telescope beam: $[$HKM99$]$ B1-bS and $[$HKM99$]$ B1-bN \citep[see][]{hir99}.}
	\tablenotetext{c}{The CB130-1 field includes 2 candidate FHCs within the telescope beam: $[$KED2011$]$ CB130-1-IRS1 and $[$KED2011$]$ CB130-1-IRS2.}
\tablerefs{(1) \cite{pin11}; (2) \cite{che10}; (3) \cite{eno10}; (4) \cite{dun11}; (5) \cite{pez12}; (6) \cite{hua13}; (7) \cite{hir14}; (8) \cite{pal14}; (9) \cite{muz03}; (10) \cite{luh99}; (11) \cite{pre01}; (12) \cite{kle03}; (13) \cite{moh05}; (14) \cite{pal12}; (15) \cite{mor15}; (16) \cite{bar09}; (17) \cite{muz05}; (18) \cite{sch06}; (19) \cite{mot01}; (20) \cite{whi04}; (21) \cite{and08}; (22) \cite{shi00}; (23) \cite{you03}; (24) \cite{taf04}; (25) \cite{fro05}; (26) \cite{cod97}; (27) \cite{and99}; (28) \cite{bel02}; (29) \cite{dun06}; (30) \cite{you06}; (31) \cite{bou06}; (32) \cite{pha11}; (33) \cite{ung82}; (34) \cite{har99}; (35) \cite{wan01}; (36) \cite{gue07}; (37) \cite{sch10}; (38) \cite{luh07}; (39) \cite{mar01}; (40) \cite{pas03}; (41) \cite{ste07a}; (42) \cite{ste07b}; (43) \cite{bou08}; (44) \cite{kim11}; (45) \cite{lee09}; (46) \cite{lee13}; (47) \cite{kau08}; (48) \cite{dun10}; (49) \cite{sch12}; (50) \cite{kau11}; (51) \cite{bou05}; (52) \cite{mah11}.}
\end{deluxetable*}

\subsection{Very Large Array}

Prompted by a detection in our single-dish observations, we observed a field centered at source L1448 IRS 2E with the
Karl G. Jansky Very Large Array (VLA) of the National Radio Astronomy Observatory on 2015 October 14 (project 15B-366). The array was in its D configuration. A baseband centered at 22 GHz was observed with an 8-bit sampler (for a high spectral resolution). Within this baseband, a subband of 8 MHz was used to observe the $6_{16}-5_{25}$ water maser line and Doppler tracking was applied, centering the subband at a frequency corresponding to $V_{\rm LSRK} = 0$  km s$^{-1}$. This subband was sampled over 2048  channels, thus providing a velocity coverage and resolution of 108 and 0.052 km s$^{-1}$, respectively. Seven subbands of 128 MHz each were also observed in this 8-bit baseband, covering a total of 896 MHz for radio continuum. In addition, two basebands, centered at 23.45 and 25.45 were also observed with 3-bit samplers. Each of these basebands was covered by 16 subbands of 128 MHz. Therefore, the combined coverage for radio continuum data provided by the three observed basebands was 3.67 GHz, centered at 23.98 GHz.

Sources J0319+4130 (3C 84), J0137+331 (3C 48), and J0336+3218 (4C 32.14) were used to calibrate the spectral bandpass response, the absolute flux scale, and the complex gain (phase and amplitude), respectively. Calibration, imaging, and further processing was carried out with the Common Astronomy Software Applications (CASA) package. All maps were obtained with a Briggs weighting of visibilities (with robust=0.5 in task {\sc clean} of CASA), and deconvolved with the CLEAN algorithm. 
In the case of radio continuum emission, multifrequecy synthesis was used to obtain the maps. The resulting FWHM of the synthesized beams were $3.4''\times 3.1''$ (p.a. $=90^\circ$) for the water maser images, and $3.1''\times 2.7''$ (p.a. $=79^\circ$) for the continuum { images}. The final images have been corrected by the primary beam response of the VLA. The absolute positional accuracy of our interferometric observations was $\simeq 0.3''$.

\section{Results}

\begin{figure}
	\plotone{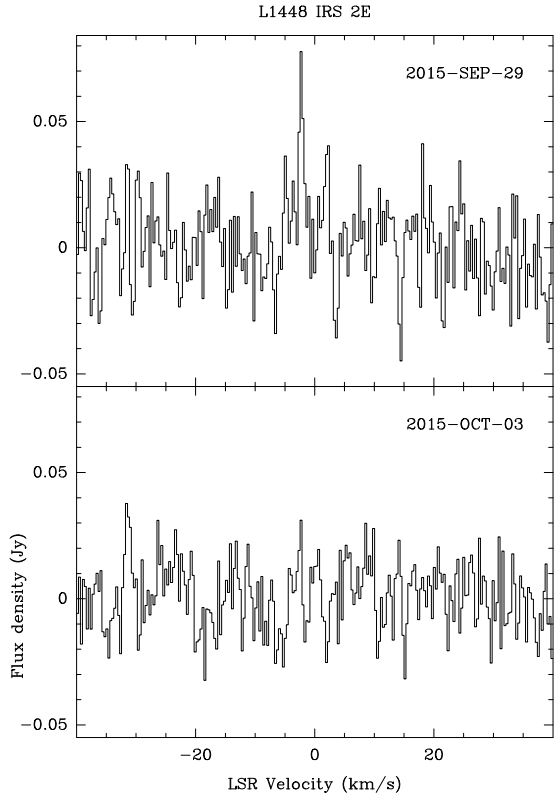}
	\caption{Water maser spectra towards the L1448 IRS 2E field obtained at the Effelsberg Radio Telescope.}
	\label{fig:spec}
\end{figure}

In our single-dish observations, we only detected maser emission toward the position of L1448 IRS 2E (Fig. \ref{fig:spec}) on 2015 September 29. It only shows one clear spectral component of flux density $80\pm 30$ mJy at $V_{\rm LSRK}=-2.3\pm 0.3$ km s$^{-1}$. This source has been classified as a FHC by \cite{che10}.

\begin{figure}
	\plotone{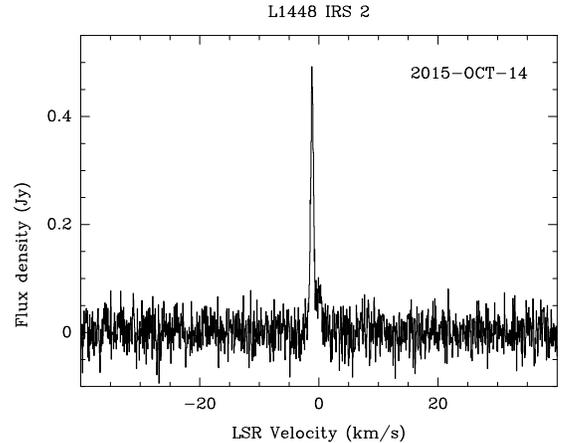}
	\caption{Water maser spectrum toward  L1448 IRS 2, obtained with the VLA.}
	\label{fig:spec_vla}
\end{figure}

However, our follow-up VLA observations shows that the maser emission is located at RA(J2000) = $03^h 25^m 22\fs 348$, Dec(J2000) = $+30^\circ 45' 13\farcs 11$. 
This position is $42''$ away from L1448 IRS 2E and { is coincident with} L1448 IRS 2 instead.
The VLA spectrum is presented in Fig. \ref{fig:spec_vla}. It shows a spectral component of flux density $460\pm 30$ mJy at $-1.19\pm 0.05$ km s$^{-1}$. 
{ The peak velocity of the emission is redshifted by $\simeq 1.1$ km s$^{-1}$ with respect to the peak measured in our single-dish spectrum}. 
{ We think that the most likely situation is that the emission detected both with Effelsberg and the VLA arises from L1448 IRS 2.
Assuming this is the case, we estimated the expected flux density in our single-dish observations, taking into account the FWHM of the Effelsberg beam ($39''$), and the distance between L1448 IRS 2 and our observed position ($42''$). For a source flux density of 460 mJy (Fig. \ref{fig:spec_vla}), the flux density in the Effelsberg spectrum should have been $\simeq 18$ mJy. This is consistent with the non-detection on our second single-dish spectrum, taken on 2015-10-03 (Fig. \ref{fig:spec}). Our spectrum on 2015-09-29 is a factor of 4 brighter than our expectation, and it would indicate a decrease in the water maser emission over the time span of 15 days covered by our data.} 
However, { although unlikely,} we cannot discard that L1448 IRS 2E (or other nearby source) was the emitting source of the spectrum in Fig. \ref{fig:spec}) and it faded over the following days. In any case, { what is relevant for our work is that} there is no confirmation that the FHC L1448 IRS 2E is a water-maser emitter. In these VLA observations, the $3\sigma$ upper limit for water maser emission toward L1448 IRS 2E is 14 mJy, after Hanning-smoothing the VLA data up to a velocity resolution of 0.21 km s$^{-1}$.

On the other hand L1448 IRS 2 is a Class 0 protostar with a luminosity $\simeq 3.6-5.2$~$L_\odot$ \citep{oli99,tob15}. Ours is the first reported detection of a water maser in this object.  \cite{fur03} and \cite{sun07} reported non-detections toward this source, with $1\sigma$ rms levels of 80 and 170 mJy, respectively.

\begin{figure}
	\plotone{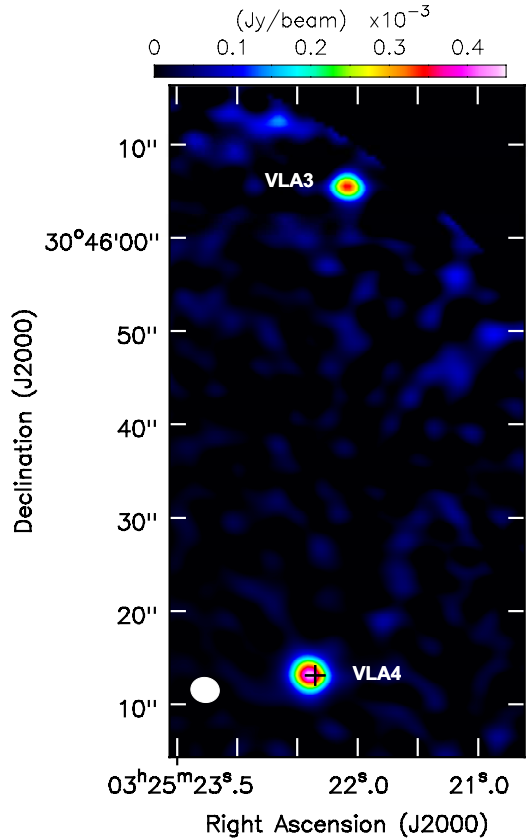}
	\caption{Image of the radio continuum sources at 1.25 cm in the L1448 IRS 2E field. The black cross near the southern source marks the position of the detected water maser. The white ellipse at the bottom left corner represents the FWHM of the synthesized beam { of the continuum image}. The detected sources are labeled following the nomenclature of \cite{ang02}.}
	\label{fig:map_vla}
\end{figure}

We detected two unresolved radio continuum sources with the VLA (Fig. \ref{fig:map_vla}), which correspond to sources VLA 3 and VLA 4 detected by \cite{ang02} at 3.6 cm. We detected source VLA 3 (the northern one in the image) at RA(J2000) $= 03^h25^m22\fs 084$, Dec(J2000) $= +30^\circ 46' 05\farcs 46$ with a flux density $S(23.98 {\rm GHz}) = 0.43\pm 0.05$ mJy. Its flux density is higher at lower frequencies \citep[e.g., $0.89\pm 0.03$ mJy at 3.6 cm,][]{ang02}, confirming the presence of non-thermal emission, which suggests that this is a background source \citep{ang02}, unrelated with this star-forming region.

The southern source is located at RA(J2000) $= 03^h25^m22\fs 400$, Dec(J2000) $=+30^\circ 45'13\farcs 22$ and it has a flux density $S(23.98 {\rm GHz}) = 0.499\pm 0.018$ mJy. This is source VLA 4 \citep[$0.34\pm 0.03$ at 3.6 cm,][]{ang02}, which is the radio counterpart of the infrared source L1448 IRS 2. Fig. \ref{fig:map_vla} clearly shows that the water maser emission is associated { with} this radio source. \cite{tob15} resolved the radio emission from L1448 IRS 2 into a binary, with a separation between components of $\simeq 0.751\pm 0.004$ arcsec. We cannot resolve this binary in our radio continuum maps, but we note that the position of the maser emission is closer to component B. This is the weaker radio source of the two components, but it has a steeper radio spectral index, which would indicate that its free-free emission is optically thicker. It could be the younger of the two components and its association with water maser emission indicates that it is undergoing energetic mass-loss.

%
%

\section{Discussion}

We did not detect any water maser emission in our sample of low luminosity objects. This is in contrast with the high detection rates  \citep[$\sim 40$\% in Class 0 YSOs,][]{fur01} in objects with higher masses. An important question is whether shocks related to mass-loss processes in these extremely low luminosity objects are energetic enough to produce water maser emission. The observations presented here are sensitive enough to impose significant constraints on our sample, to determine whether there is a lower limit on the stellar luminosity of sources capable to excite this type of emission.

Previous water maser surveys established an empirical relationship between the stellar luminosity and the water maser luminosity \citep[e.g.,][]{bra03,shi07}. For instance, \cite{shi07} give $L_{\rm H_2O} = 3\times 10^{-9} L_{\rm bol}^{0.94}$. A correlation between the maximum velocity spread of the spectral components of the maser emission and the maser luminosity has also been found \citep{bra03}. 
Assuming a 3-$\sigma$ upper limit to the flux density of the water maser emission in our sample of $\simeq 60$ mJy 
(since { the rms was typically} $\simeq 20$ mJy, Table \ref{tab:observed}), and an upper limit to the velocity spread of masers of 
$\simeq 1$ km s$^{-1}$ \citep[similar to that found in source GF 9-2, with 0.3 $L_\odot$,][]{fur03}, we obtain upper limits for the maser luminosity of our sample of $7\times 10^{-11}$ $L_\odot$ for sources in Perseus \citep[assuming a distance of 232 pc,][]{hir11} and $3\times 10^{-11}$ $L_\odot$ for those in Taurus  \citep[distance 137 pc,][]{tor07}. Most of our sources are in those two complexes (those with right ascension between 3 and 5 hours).

\begin{figure}
	\plotone{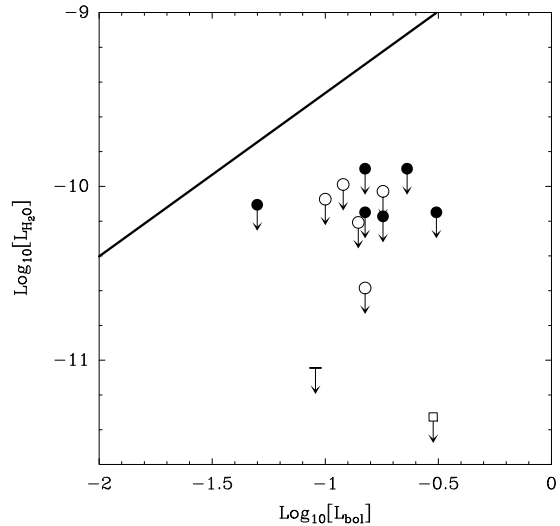}
	\caption{Upper limits of the logarithm of water maser luminosity in FHC and Class-0 objects in our sample, versus the logarithm of bolometric luminosities. The luminosities in both axes are measured in units of solar luminosities. Filled circles represent FHCs, and open circles represent Class-0 objects (both YSOs and BDs). We also plotted the upper limits for other two objects with $L_{\rm bol} < 1$ L$_\odot$: GF 9-2 (open square; Manjarrez et al. in preparation) and L1014-IRS     \citep[horizontal bar;][]{shi07}, assuming a distance of 200 pc in both cases. The solid line represents the expected maser luminosities from the empirical relationship by \citet{shi07}. Sources with only upper limits to their bolometric luminosities in Table \ref{tab:observed} are not included, since they would not constrain the luminosity relationship.}
	\label{fig:correlation}
\end{figure}

If the empirical relationship by \cite{shi07} stands for the sources in our sample, our observations should have enough sensitivity to detect water masers from sources with $L_{\rm bol} \ga 10^{-2}$ $L_\odot$ in Perseus and Taurus. This is illustrated in Fig. \ref{fig:correlation}, where we show the upper limits obtained in our observations for the water maser luminosity in FHC and Class-0 YSOs and BDs, together with the extrapolation to low luminosities of the empirical correlation by \cite{shi07}. We also show in this figure the upper limits to maser luminosity for other two sources with $L_{\rm bol} < 1$ L$_\odot$ \citep[L1014-IRS and GF 9-2;][Manjarrez et al. in preparation]{shi07}. Although this empirical relationship has a large scatter, the upper limits in luminosity for all sources fall systematically below the empirical prediction by around one order of magnitude. The same happens with Class I and II sources in our sample, but we did not plot them in this figure because these more evolved objects have a much lower probability of pumping maser emission \citep{fur01}.

{ If the empirical relationship between stellar and maser luminosities is still valid at $L_{\rm bol}<1$ L$_\odot$ and the detection rates were similar to the ones found in other low-mass Class 0 protostars ($\simeq 40$\%),  we should have detected some  water masers. So either premise is not correct}. { The first possibility (that the empirical relationship between stellar and maser luminosities does not hold at the lowest end of the mass spectrum) would mean that the relationship is steeper for the lowest-luminosity objects, and the maser luminosities in objects $< 1$ L$_\odot$ are lower than predicted by the empirical relationship \citep[see Fig. \ref{fig:correlation} in this paper and Fig. 3 in][]{shi07}. It is even possible that} there is a strict threshold in luminosity below which, we cannot detect masers. { The second possibility} is that the relationship does stand at low luminosities, but the detection rates of masers are significantly lower than the values obtained in higher-luminosity Class 0 protostars, and even though the maser emission from these sources may { eventually} be above our detection threshold, it is excited during very short periods of time and we missed them. With our data it is not possible to discern between these possibilities. A further monitoring of these sources in our sample would help us to determine whether our failing to detect maser emission is due to a significant decrease of detection rates.

If a luminosity threshold exists, its value is still uncertain. As mentioned above, the lowest-luminosity object that has been reported to harbor a water maser is GF 9-2 \citep[L$_{\rm bol} \simeq 0.3$ L$_\odot$][]{fur03}. However, this detection was obtained with the Nobeyama single-dish radio telescope { in 2000 April}, and has never been confirmed with interferometric observations.  VLA observations { carried out in 2008 August} by Manjarrez et al (in preparation) failed to detect { any water maser emission within $\simeq 1'$ of GF 9-2}, with a 3$\sigma$ upper limit of 5 mJy, a factor $\sim 75$ lower than the flux density of reported by \cite{fur03}. { Since the beam size of the Nobeyama telescope at 22 GHz ($75''$) is smaller than the VLA primary beam ($\simeq 2'$), we can be certain the emission detected in 2000 faded away by 2008. While it is certainly possible that GF 9-2 was the water-maser emitting source in 2000, we cannot discard that the emission arose from another nearby source, as in the case of L1448 IRS 2E in this paper.}
To our knowledge, the lowest luminosity YSO in which water maser emission has been confirmed with interferometric observations is  VLA 1623 \citep{fur03}, with a bolometric luminosity of $\simeq 1$ L$_\odot$. 

\section{Conclusions}

We have carried out a sensitive single-dish survey (using the Effelsberg Radio Telescope) for water maser emission toward a sample of low-luminosity objects, comprising 20 young BDs, 7 FHCs, 6 VeLLOs, and 11 additional YSOs with L$_{\rm bol} < 0.4$ L$_\odot$ and/or $M<0.15$ M$_\odot$. We only detected water maser emission when pointing towards L1448 IRS 2E, a FHC. However, follow-up interferometric observations with the VLA did not confirm the presence of water maser emission associated with this source. Instead, we found water maser emission associated with the Class-0 YSO L1448 IRS 2. This is the first reported detection of water maser emission toward this source.

If we extrapolate to low luminosities the known correlation between bolometric luminosity and water maser luminosities, established for sources with $L_{\rm bol} \ge 1$ L$_\odot$, we found that the expected water maser luminosities are one order of magnitude higher than the upper limits obtained in our survey for FHC and Class-0 objects. Possible explanations are that  the $L_{\rm H_2O}$ vs $L_{\rm bol}$ relationship does not extend to the lower-end of the sub(stellar) mass spectrum (either its slope is steeper at low luminosities or there is a luminosity threshold below which water maser cannot be excited), or that the detection rates of water maser emission is lower than the values obtained for higher-luminosity Class 0 YSOs { ($\simeq$40\%)}.

\acknowledgments

This paper is partly based on observations with the 100-m telescope of the MPIfR (Max-Planck-Institut f\"ur Radioastronomie) at Effelsberg. The National Radio Astronomy Observatory is a facility of the National Science Foundation operated under cooperative agreement by Associated Universities, Inc. The research leading to these results has received funding from the European Commission Seventh Framework Programme (FP/2007-2013) under grant agreement No 283393 (RadioNet3). JFG is supported by MINECO (Spain) grant AYA2014-57369-C3-3 (co-funded by FEDER),  and by MECD (Spain), under the mobility program for senior scientists at foreign universities and research centers. A.P. acknowledges financial support from UNAM-DGAPA-PAPIIT IA102815 grant (Mexico). L.U. acknowledges support from PRODEP (Mexico). DB is supported by MINECO grant ESP2015-65712-C5-1-R.

\facilities{Effelsberg, VLA}
\software{GILDAS, CASA}

\end{document}